\documentclass[onecolumn,showpacs,preprintnumbers,amsmath,amssymb,superscriptaddress]{revtex4}

\usepackage{amsbsy}
\usepackage{graphicx}

\begin{document}

\title{Shape-dependent Depinning of a Domain Wall by a Magnetic Field and a Spin-Polarized Current}

\author{N. Logoboy}

\email{logoboy@phys.huji.ac.il}

\affiliation {Institute of Superconductivity, Department of Physics,
Bar-Ilan University, 52900 Ramat Gan, Israel}

\affiliation{The Racah Institute of Physics, The Hebrew University
of Jerusalem, 91904 Jerusalem, Israel}

\author{B. Shapiro}

\affiliation {Institute of Superconductivity, Department of Physics,
Bar-Ilan University, 52900 Ramat Gan, Israel}

\author{E. B. Sonin}

\affiliation{The Racah Institute of Physics, The Hebrew University
of Jerusalem, 91904 Jerusalem, Israel}

\date{\today}

\begin{abstract}

The effect of sample shape on the depinning of the domain wall (DW)
driven by an applied magnetic field or a spin-polarized current is
studied theoretically. The shape effect resulting from the
modulation of the sample width (geometric pinning) can essentially
affects the DW depinning.  We found a good agreement between the
ratios of the critical values of the magnetic field and the
spin-polarized current predicted by the theory and measured in the
experiment.
\end{abstract}

\pacs{ 72.30.+q, 75.60.Ch, 75.70.Ak, 85.75.-d}

\maketitle

\section{\label{sec:Introduction}Introduction}

During the last years an immense interest in current-driven domain
wall (DW) motion in thin magnetic films, nanotubes and point
contacts has been initialized by possible applications in spintronic
device technology (see, e. g. \cite{Marrows}, \cite{Thomas} and
references therein). These devices are expected to be highly efficient,
fast and consuming less energy. They possess such important features
as non-volatility, portability and capability of simultaneous data
storage and processing. A manipulation of magnetization by
spin-polarized current predicted by Berger for non-uniform
ferromagnets \cite{Berger} and Slonczewski for multi-layered
ferromagnetic structures \cite{Slonczewski1} has attracted the
attention of physical community during last decade and gained
further development in \cite{Li}-\cite{Klein2}.

The problem of DW motion under another driving force, applied
magnetic field, was addressed in late 70s in connection to possible
application in memory storage devices (see, e. g.
\cite{Malozemoff}). In pioneering work of Schryer and Walker
\cite{Schryer} it was shown the existence of an instability in the
laminar movement of the DW. These early results can actually serve
as a reference for the studies of spin-polarized current-driven DW
motion. In particular, it is well established now that for DW driven
by magnetic field (or spin-polarized current, or both) there exists
a critical value of driving parameter(s) which corresponds to
maximum velocity of the DW (Walker velocity \cite{Schryer}). Thus,
the DW dynamics includes two distinct regimes, so-called subcritical
(below the Walker breakdown) and supercritical (above the Walker
breakdown) ones \cite{Marrows}, \cite{Thiaville}). Below the Walker
limit the overdamped transient response of the system to applied
magnetic field follows by a steady state response, while above the
Walker limit the DW dynamics is oscillatory with non-zero average
velocity.

In case of field-driven DW motion the exact stationary wave solution
in subcritical regime was obtained more than half-century ago by
Walker \cite{Walker}. Analytical results of Bourne \cite{Bourne}
reproduced a velocity profile in full range of magnetic field
confirming numerical simulations of Slonczewski in supercritical
regime \cite{Slonczewski2}.

Modern fabricated logic elements based on manipulation of the DWs
are represented by a complex networks of nanowires, which can form
three-dimensional memory storage structure, e. g., a racetrack device
that comprises an array of magnetic nanowires arranged horizontally
or vertically on a silicon chip \cite{Parkin}. The spacing between
consecutive DWs is controlled by pinning sites fabricated by
patterning notches along the edges of the track or modulating the
track's size and material properties. Being a necessary component
of the logic elements, the pinning sites define the bit length and
provide the DWs stability against external perturbations, such as
thermal fluctuations or stray magnetic fields from nearby
racetracks.  The variation of the nanowire geometry creates the
pinning potential for the DW. The knowledge of the behavior of the
DW in artificially created structural defects and constrictions is
extremely important for producing reliable memory devices. In spite
of a growing number of experimental studies which gain insight into
the properties of the DWs pinned by artificial defects (see, e. g.
\cite{Klein1}, \cite{Bogart}-\cite{Ahn}), a detailed understanding
of the role of the shape effects on the dynamics of the DW driven by
magnetic field and (or) spin-polarized current is still lacking.

In this paper we study shape-dependent effects on the properties
of DW confined in potential wells created by the
defects in the bulk
and by the variation of the sample shape (geometrical pinning). We analyze the difference in
the depinning of the DW driven by applied magnetic field, from one
side, and spin-polarized current, from another. We provide a plausible
explanation of recent experimental data on manipulation of DW in
constricted stripes of $\mathrm{SrRuO_{3}}$ \cite{Klein1}.

\section{\label{sec:GBE}Geometry and Basic Equations}

The DW displacement under driving magnetic field and spin-polarized
current can be adequately described by Landau-Lifschitz-Gilbert
(LLG) equations complemented by a spin-transfer torque, $\tau_{J}$:
\begin{equation} \label{eq:LLG}
\partial_{t}\mathbf{m}=-\gamma [\mathbf{m}\times \mathbf{H}_{e}]+\alpha [\mathbf{m}\times
\partial_{t}\mathbf{m}]+\mathbf{\tau}_{J}, \quad
\end{equation}
where $\gamma$ is the gyromagnetic ratio, $\alpha$ is the dissipation parameter, $\mathbf{H}_{e}=-\delta F/\delta \mathbf{M}$ is effective
magnetic field, $F$ is free energy density of the ferromagnet, and
$\mathbf{m}=\mathbf{M}/M_{s}$ is a unit vector in the direction of
the magnetization $\mathbf{M}$ ($M_{s}$ is the saturation
magnetization).

Despite different approaches in calculation of the spin-transfer
torque $\mathbf{\tau}_{J}$, there is a consensus about the existence
of adiabatic and non-adiabatic terms (or $\beta$-term
\cite{Tatara1}) that contribute to the spin-transfer torque
\begin{equation} \label{eq:STT}
\mathbf{\tau}_{J}= -(\mathbf{U\cdot\triangledown})\mathbf{m}+\beta
\mathbf{m}\times(\mathbf{U\cdot\triangledown})\mathbf{m},
\end{equation}
where $\mathbf{U}=U\mathbf{n}$, $U=\mu_{B}JP/eM_{s}$ ($\mu_{B}$ is
Bohr magneton, $e$ is the elementary charge,
$J=\mathcal{I}/\mathcal{A}$ is the current $\it{density}$,
$\mathcal{I}$ is a value of the current, $\mathcal{A}$ is a
cross-section area of the sample and $P$ is the spin polarization),
$\mathbf{n}$ is unit vector in the direction of the current.
Parameter $\beta$ is a ratio between the values of non-adiabatic and
adiabatic torques.

\begin{figure}[t]
  \includegraphics[width=0.4\textwidth]{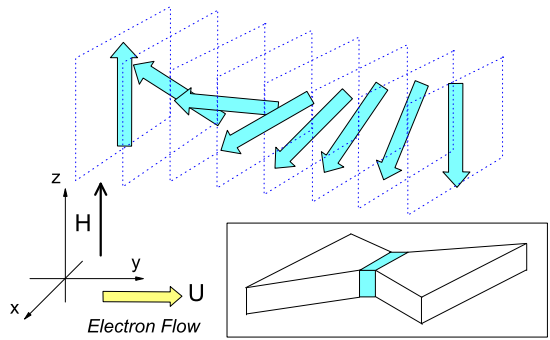}
\caption{(color online) $180^{0}$ Bloch domain wall. The
magnetization in the domains are parallel and antiparallel along the
$z$ axis (the easy axis). Inside the domain wall the magnetization
rotates in the $xz$ plane. The low panel shows an example of
constricted sample with the DW located in the constriction.}
\label{Fig_1}
\end{figure}

Let us consider the $180^0$ DW of Bloch type in a constricted
plate-like sample with variable size $L_{x}=L_{x}(y)\equiv
L_{0}[1+\mathcal{G}(y)]$ along $x-$direction and constant dimensions
$L_{y}$ and $L_{z}$ along $y-$ and $z-$axis as shown in
Fig.~\ref{Fig_1}. The function $\mathcal{G}=\mathcal{G}(y)$
describes the change of the sample shape. In absence of
constriction $\mathcal{G}=0$ and $L_{x}=L_{0}$. The DW width,
$\Delta$, is much less than the width of the plate, $L_{z}$, i.e.
$\Delta\ll L_{z}$. The surface of the sample is
parallel to the $xy$-plane, and the domain wall, being
parallel to the $xz$-plane, separates two domains with magnetization
$M(y)$ along the $+z$ or $-z$ direction and is located initially in
the constriction at $y=$ 0.

The free energy functional of the ferromagnet is
$\mathcal{F}=\int_{V} \mathbf{dr}~F$ where the free energy density
$F=F[\mathbf{m}(\mathbf{r}),
\partial \mathbf{m}(\mathbf{r})/\partial x_{i}]$ is defined by
\begin{eqnarray} \label{eq:Free Energy Density}
F=K\left[ -m^{2}_{z}+q~m^{2}_{y}+\Delta^{2}_{0}(\partial_{i}
m_{j})(\partial_{i}
 m_{j})\right]-HM_{s}m_{z}~.
\end{eqnarray}
Here, the first and second terms in square brackets describe the
magnetic crystallographic anisotropy, the third term is the exchange
energy and the last term is the Zeeman energy in an external magnetic field
$\mathbf {H}|| \mathbf{z}$. In Eq.~(\ref{eq:Free Energy Density})
$K>$0 is the parameter of the easy-axis crystallographic anisotropy,
the ratio $q=K_{\perp}/2K>$0 determines a joint effect of orthorhombic anisotropy
and the magnetostatic energy, and  $\Delta_{0}$ is a half-width of the DW at  rest, which
characterizes the stiffness of the spin system. The
choice of the sign $q>$0 implies  alignment of the magnetization
$\mathbf{M}$ in the $xz$-plane. Thus, in the absence of the driving
force the formation of DW of Bloch type with the rotation of the
magnetization in $xz$-plane is energetically more favorable.

The magnetization $\mathbf{m}$ can be expressed in polar
coordinates, $ \mathbf{m} = (\sin\theta \cos\phi, \sin\theta
\sin\phi, \cos\theta)$. In the absence of driving forces
($\mathbf{H, U}=$0) the ground state of the system is defined by
minimization of the free energy with respect to azimuthal, $\phi$,
and polar, $\theta$, angles, i.e., $\delta \mathcal{F}/\delta
\phi=0$ and $\delta \mathcal{F}/\delta \theta=0$. This yields the
well-known structure of Bloch DW located at the center of the sample
constriction ($y=$0): $\theta_{0}=2\tan^{-1} \exp{(y/\Delta_{0})}$
and $\phi_{0}=0$.

The standard approach in statics and dynamics of the DW is to use
the Slonczewski equations for two canonically conjugated variables,
the coordinate of the center of the DW  $\eta$, and azimuthal angle
$\phi$ (see, e. g. \cite{Malozemoff}), which are independent of the
coordinate $y$. For a sample with variable cross-section area
$\mathcal{A}(y)=\mathcal{A}_{0}[1+\mathcal{G}(y)]$
($\mathcal{A}_{0}=L_{z}L_{0}$ is the area of the cross-section at
$y=$ 0, and $\mathcal{G}(\eta)$ is a shape function  dependent on
the geometry of the sample) these equations are:
\begin{equation}\label{eq:DWE1}
(\frac{d\phi }{dt}+\frac{\alpha}{\Delta} \frac{d\eta
}{dt})[1+\mathcal{G}(\eta)]=-\frac{\gamma}{2M_{s}}\frac{\delta\sigma}{\delta\eta}+\beta\frac{U_{0}}{\Delta}~,
\quad (\frac{d\eta }{dt}-\alpha\Delta\frac{d\phi
}{dt})[1+\mathcal{G}(\eta)]=\frac{\gamma}{2M_{s}}\frac{\delta\sigma}{\delta\phi}+
U_{0}~,
\end{equation}
where $U_{0}=[1+\mathcal{G}(y)]U(y)$,
$\Delta=\Delta_{0}(1+q\sin^{2}\phi)^{-1/2}$ is the effective DW
width, and $\sigma(\eta, \phi) $ is the surface energy of the DW
determined by the integral across the whole DW width:
\begin{equation}\label{eq:DW Free Energy Density} \sigma=
\mathcal{A}_{0}    \int dy~ [1+\mathcal{G}(y)]~F[\theta_{0}(y-\eta),
\phi].
\end{equation}

The goal of the present work is not investigation of the DW motion
but determination of the depinning threshold, when such motion
becomes possible. Therefore, one may ignore the time dependence of
the variables. The second term $\propto \beta$ in the spin torque
(nonadiabatic torque), Eq. (\ref{eq:STT}), is most relevant for
depinning. This  term is related to the momentum transfer between
the polarized current and the DW.  The adiabatic torque [the first
term in Eq. (\ref{eq:STT})] causes the rotation of the DW plane
around $z-$axis relative to it equilibrium position ($xz-$plane),
but does not affect the DW depinning directly.

\section{\label{sec:Results and Discussions} Depinning of Domain Walls}

Usually, the pinning (coercivity) force on a DW originates from
randomly located defects, which create potential wells for the DW in
the sample bulk. However, key features of the bulk pinning
phenomenon can be investigated using a simpler model of a DW in a
periodic potential. The latter enters the DW surface energy, i.e.,
$\sigma \to \sigma +\sigma_{p}$ with
$\sigma_{p}=2H_{p}M_{s}[1+\mathcal{G}(\eta)]\xi f(\eta)$, where
$f(\eta+2\xi)=f(\eta)$ is a periodic function with a period 2$\xi$.
Within the period ($-\xi\le\eta\le\xi$) this function can be
approximated by a parabolic function $f(\eta)=(1/2)(\eta/\xi)^{2}$.
In addition to the pinning on the defects, there is another type of
pinning (geometrical pinning) related to the change of the sample
shape. Eventually neglecting the structure of the DW, i.e., assuming
that $\sin^{2}\theta(y)=2\Delta \delta(y)$, we obtain
\begin{equation}\label{eq:DWE2}
\beta\frac{U_{0}}{1+\mathcal{G}(\eta)}+\gamma H \Delta=\gamma
H_{a}\Delta^{2}_{0}~\frac{\partial_{\eta}
\mathcal{G}(\eta)}{1+\mathcal{G}(\eta)}+\gamma
H_{p}\xi\partial_{\eta}f(\eta)\Delta~, \quad
\frac{U_{0}}{1+\mathcal{G}(\eta)}=-\frac{q}{2}\gamma H_{a}\Delta\sin
2\phi~,
\end{equation}
where $H_{a}= 2K/M_{s}$ is a magnetic field corresponding to the
easy-axis magnetic crystallographic anisotropy of a sample.
According to Eq.~(\ref{eq:DWE2}), the modulation of the sample width
gives rise to an effective geometrical pinning of the DW. It follows
from Eqs.~(\ref{eq:DWE2}) that DW can be depinned  by action of
magnetic field, $H$, or the nonadiabatic contribution of the
spin-polarized current, $\beta\ne\mathrm{0}$.

In the following, we consider two particular cases,
$H\ne\mathrm{0}$, $U=\mathrm{0}$ and $U\ne\mathrm{0}$,
$H=\mathrm{0}$.

\subsection{\label{sec:Depinning by applied magnetic
field}Depinning by applied magnetic field: $H\ne\mathrm{0}$,
$U=\mathrm{0}$ }

In case of $H\ne$ 0 and $U=\mathrm{0}$, it follows from
Eqs.~(\ref{eq:DWE2}), that $\phi=\mathrm{0}$, $\Delta=\Delta_{0}$,
and the depinning of the DW by magnetic field is not affected by
the presence of the orthorhombic anisotropy. Thus, instead of the first equation in
(\ref{eq:DWE2}) we have
\begin{equation}\label{eq:Depinning by Magnetic Field2}
[H-H_{p}\xi\partial_{\eta}f(\eta)][1+\mathcal{G}(\eta)]-H_{a}\Delta_{0}~\partial_{\eta}
\mathcal{G}(\eta)=0~,
\end{equation}
Equation~(\ref{eq:Depinning by Magnetic Field2}) manifests the absence of a total force experienced by DW. The
driving force from the magnetic field (the term $\propto H$) and  the bulk  pinning
force confining the DW in a potential well (the term $\propto H_p$) are proportional to the
total DW area, $\sim [1+\mathcal{G}(\eta)]$, while the force
experienced by the DW in a shape-dependent pinning potential is determined by the derivative
of the shape function  $\partial_{\eta} \mathcal{G}(\eta)$
[the last term in Eq.~(\ref{eq:Depinning by Magnetic Field2})].

For simple  shape potential, which can be expanded in series
on DW displacement $\eta$, the function $\partial_{\eta}
\mathcal{G}(\eta)/[1+\mathcal{G}(\eta)]$ reaches a maximum at some
value of the parameter $\eta=\zeta$. Thus, it is insightful to
characterize the geometric pinning potential, which is responsible
for the shape effect, by the strength of the potential $H_{\zeta}$
and the characteristic distance  $\zeta$, which are analogous to the
parameters $H_{p}$ and $\xi$ of the pinning potential due to the
defects. The value of the critical magnetic field $H_{\zeta}$ is
defined by
\begin{equation}\label{eq:Shape Critical Depinning Field}
H_{\zeta}=\max\left\{H_{a}\Delta_{0}\frac{\partial_{\eta}\mathcal{G}(\eta)}{1+\mathcal{G}(\eta)}\right\}=H_{a}\Delta_{0}\frac{\partial_{\zeta}\mathcal{G}(\zeta)}{1+\mathcal{G}(\zeta)}~,
\end{equation}
while $\zeta$ is obtained from the condition of a potential extremum
$ (1+\mathcal{G})\partial^{2}_{\zeta\zeta}\mathcal{G}-(\partial_{\zeta}\mathcal{G})^{2}=$
0.
\begin{figure}[b]
  \includegraphics[width=0.4\textwidth]{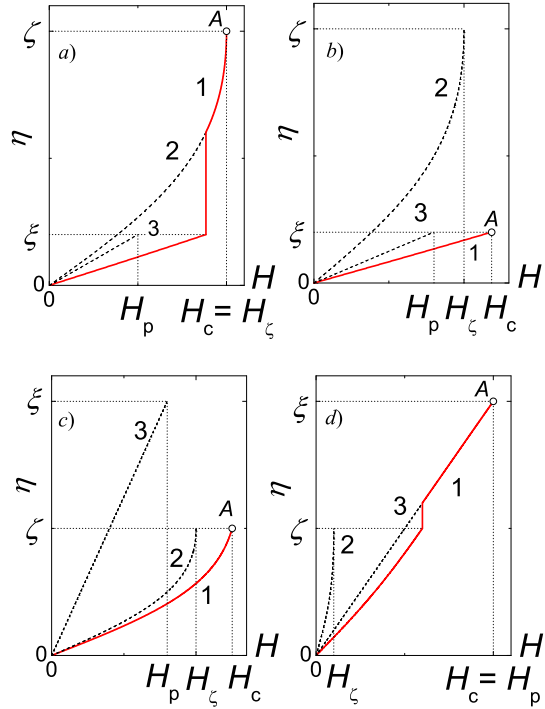}
\caption{ (Color online) The plots
$\eta=\eta(H)$, which illustrate four different scenarios for the
depinning of the DW subjected to an applied magnetic field $H$, for
$\zeta>\xi$ [(a) and (b)) and $\zeta\le\xi$ ((c) and (d)], as
described in the text. The solid curves 1 show the DW displacement $\eta$ for the joint effect of
bulk pinning  and geometrical pinning potentials. The DW displacement in the presence only of bulk pinning or geometric pinning  is shown
by dashed curves 2 and 3 respectively. The open circle
corresponds to critical point $A(\eta_{c}, H_{c})$,  where eventually depinning takes place.
The critical magnetic field $H_{c}$ and  the critical
displacement of the DW  $\eta_{c}$ are defined by
Eqs.~(\ref{eq:Depinning Field}) and (\ref{eq:Crititical
Displacement}).} \label{Fig_2}
\end{figure}

Thus, the presence of a constriction results in the change of the
DW area and appearance of the geometrical pinning, which is
independent of the distribution of  defects. In absence of
 pinning on defects ($H_{p}= 0$), the increase in the applied
magnetic field below the critical field ($H<H_{\zeta}$) causes the DW
displacement $\eta (H)$ (which is not  a linear function of
a magnetic field in general) until its depinning at $H=H_{\zeta}$.

With use of $H_{\zeta}$ and $\zeta$ both terms that contribute to
the DW pinning [see  Eq.~(\ref{eq:Depinning by Magnetic Field2})]
can be rewritten in a similar way:
\begin{equation}\label{eq:Depinning by Magnetic Field3}
H-H_{p}\frac{\eta}{\xi}\Theta(\xi-|\eta|)-H_{\zeta}\psi(\eta)\Theta(\zeta-|\eta|)=0~,
\end{equation}
where $\Theta(\xi)$ is a step-function, and the function
$\psi(\eta)$ is defined according to
\begin{equation}\label{eq:psi}
\psi(\eta)=~\frac{\partial_{\eta}\mathcal{G}(\eta)[1+\mathcal{G}(\zeta)]}{\partial_{\zeta}\mathcal{G}(\zeta)[1+\mathcal{G}(\eta)]}.
\end{equation}
Equation~(\ref{eq:Depinning by Magnetic Field3}) defines a function
$\eta=\eta(H)$ and can be solved for given geometry of the
constriction. The results of numerical calculation of the function
$\eta=\eta(H)$ in the particular case of a parabolic potential created by
the constriction when $\mathcal{G}=(\eta/\zeta)^{2}$ are shown in
Fig.~\ref{Fig_2} and reveal the general features of the DW depinning
by the applied magnetic field $H$. Figure~\ref{Fig_2} illustrates the
possibility of four different scenarios of DW depinning dependent on
the relation between $\xi$ and $\zeta$, from one side, and $H_{p}$
and $H_{\zeta}$, from another. The analysis shows that these scenarios
can be classified according to the values of two critical
parameters, namely,  the depinning magnetic field $H_{c}$ and critical
value of the DW displacement $\eta_{c}$. The value of $H_{c}$ and
$\eta_{c}$ are defined according to
\begin{equation}\label{eq:Depinning Field}
H_{c}=\left\{
      \begin{array}{ll}
       \max\{ H_{p}~; H_{\zeta}+(\zeta/\xi)H_{p}\}, & \textrm{if $\zeta\le\xi$}\\
       \max\{ H_{\zeta}~; H_{p}+H_{a}\Delta_{0}\partial_{\xi}\mathcal{G}(\xi)/[1+\mathcal{G}(\xi)], & \textrm{if $\zeta > \xi$}\\
      \end{array} \right.
\end{equation}
and
\begin{equation}\label{eq:Crititical Displacement}
\eta_{c}=\left\{
      \begin{array}{ll}
       \xi, & \textrm{if $H_{c}=H_{p}$, $H_{p}+H_{a}\Delta_{0}\partial_{\xi}\mathcal{G}(\xi)/[1+\mathcal{G}(\xi)]$ }\\
       \zeta, & \textrm{if $H_{c}=H_{\zeta}$, $H_{\zeta}+(\zeta/\xi)H_{p}$ }\\
      \end{array} \right.~,
\end{equation}
where we assume that the maximum width of the constriction along
$y$-axis exceeds the critical distance $\eta_{c}$. For $H\ge H_{c}$
($\eta\ge\eta_{c}$) the pinning potential cannot stop motion of the DW.

\begin{figure}[b]
\includegraphics[width=0.4\textwidth]{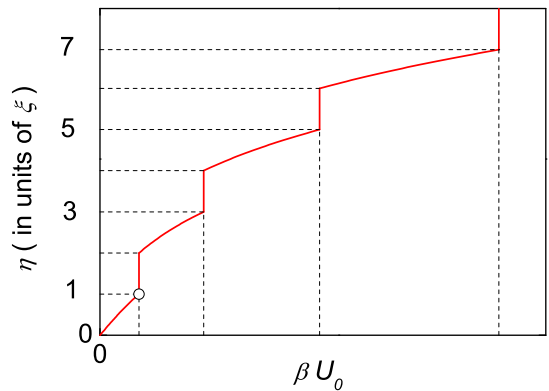}
\caption{ (Color online) The plots $\eta=\eta(\beta U_{0})$ of the DW confined in
potential wells created by artificially fabricated constriction and
bulk  pinning centers.
The jumps of the function $\eta=\eta(\beta U_{0})$ are the result of
the DW depinning from local volume defects within periods of the
potential landscape. The values of the current-dependent parameter $\beta U_{0}$  required to
depin the wall from the local potential are shown by the vertical
dashed lines. The increase in the driving force results in a
step-by-step drift of the DW. In calculations we use a parabolic
shape function $\mathcal{G}=(\eta/\zeta)^{2}$. Open circle
corresponds to the critical parameters for a sample without
constriction. Note the absence of the depinning critical point for the constricted
sample.} \label{Fig_3}
\end{figure}

\subsection{\label{sec:Depinning by spin-polarized
current}Depinning by a spin-polarized current: $U\ne\mathrm{0}$,
$H=\mathrm{0}$}

In the case of $U\ne\mathrm{0}$, $H=\mathrm{0}$, instead of
Eqs.~(\ref{eq:DWE2}) we have another pair of equations:
\begin{eqnarray}
\beta U_{0}-\gamma H_{a}\Delta^{2}_{0}~\partial_{\eta}
\mathcal{G}(\eta)-[1+\mathcal{G}(\eta)]\gamma H_{p}\Delta\xi
\partial_{\eta} f(\eta) =0~, \label{eq:Depinning by Current 1} \\
\frac{U_{0}}{1+\mathcal{G}(\eta)}=-\frac{q}{2}\gamma H_{a}\Delta\sin
2\phi~. \label{eq:Depinning by Current Angle 1}\qquad \qquad \quad
\end{eqnarray}
Equation~(\ref{eq:Depinning by Current 1}) manifests the balance of forces on the DW: the forces from the spin-polarized current ($\sim \beta$) and the  geometrical pinning ($\sim H_{a}$),  and
the bulk pinning force ($\sim H_{p}$).
Equation~(\ref{eq:Depinning by Current Angle 1}) defines the function
$\phi=\phi(\eta, U_{0})$:
\begin{equation}\label{eq:Depinning by Current Angle 2}
\sin
\phi=-~\mathcal{U}\{\frac{1}{2}(1-q~\mathcal{U}^{2})+[\frac{1}{4}(1-q~\mathcal{U}^{2})^{2}-\mathcal{U}^{2}]^{1/2}\}^{-1/2}
~,
\end{equation}
where $\mathcal{U}=U_{0}/2q\gamma
H_{a}\Delta_{0}[1+\mathcal{G}(\eta)]$. Substituting
Eq.~(\ref{eq:Depinning by Current Angle 2}) into
Eq.~(\ref{eq:Depinning by Current 1}) one can calculate the function
$\eta=\eta(U_{0})$. For the sake of simplicity we neglect the change of the DW
width assuming that $\Delta\approx\Delta_{0}$, which is true if
$\phi\ll$ 1 ($\mathcal{U}\ll$ 1).  In this case Eq.~(\ref{eq:Depinning by Current 1}) can be
rewritten in a following way:
\begin{equation}\label{eq:Depinning by Current 2}
\frac{\beta U_{0}}{\gamma\Delta_{0}}-
H_{a}\Delta_{0}~\partial_{\eta}
\mathcal{G}(\eta)-[1+\mathcal{G}(\eta)] H_{p}\xi
\partial_{\eta} f(\eta)=0~.
\end{equation}
It follows from Eq.~(\ref{eq:Depinning by Current 2}), that in an
unconstricted sample ($\mathcal{G}=$0) a DW can be depinned and
participate in a steady-state motion when the
the current exceeds the critical value defined by the
equation $\beta U^{c}_{0}/\gamma\Delta_{0}=\max{H_{p}\xi
\partial_{\eta} f(\eta)}$.

The condition for DW geometric depinning by the current is more severe than by the magnetic field. This because the pressure on the DW from the constant magnetic field (the force per unit area of the DW) does not depend on the DW area growing with the DW displacement, whereas the pressure from the constant current is determined by  the current density, which is inversely proportional to the DW area. According to Eq.~(\ref{eq:Depinning by Current 2}) at $H_p=0$, the critical   value of a current is determined by the threshold $\beta U^{(g)}_{0}=\gamma
H_{a}\Delta^{2}_{0}\max\{\partial_{\eta} \mathcal{G}(\eta)\}$ above which
the DW can overcome the geometrical pinning.  For a parabolic shape function $\mathcal{G}=(\eta/\zeta)^{2}$, the function $\partial_{\eta} \mathcal{G}(\eta)$ is unbound,  and DW depinning from the geometric pinning potential is impossible. Let us consider the DW behavior in this case.

 We assume that a
single DW is initially located in the constricted area ($\eta=$ 0).
The increase of the current density above its critical value at a
given position of the DW results in the drift of the DW from the
constriction into expansion part of the sample. Such a drift is
accompanied by the increase in the DW area followed by the decrease
of the current $\it{density}$ (the current does not depend on the
location of the DW) below its critical value. As a result, the DW
will be eventually pinned in a new position by an array of the
nearest defects. The further increase in the current  results in increase in a driving force and a
displacement of the DW into a new position where the driving force
is balanced by the increase in the pinning force. At this new
position the DW is stuck again till the next increase of the
current. Actually, in the presence of bulk pinning centers the
displacement of the DW under spin-polarized current  is
characterized by a step-by-step drift over an array of bulk defects.

Equation~(\ref{eq:Depinning by Current 2}) can be solved numerically for
given functions $\mathcal{G}(\eta)$ and $f(\eta)$. The results of
numerical calculation of the function $\eta=\eta(\beta U_{0})$ for a
parabolic shape function $\mathcal{G}=(\eta/\zeta)^{2}$ are present
in Fig.~\ref{Fig_3} which illustrates the absence of the critical
point on the curve $\eta=\eta(\beta U_{0})$ ($\eta \to \infty$ when
$\beta U_{0} \to \infty$), contrary to the case of depinning of
a DW driven by a magnetic field $H$ (see Fig.~\ref{Fig_2}).

\subsection{\label{sec:Experiment} Domain wall in a constricted sample of $\mathrm{SrRuO_{3}}$ and comparison with experiment}

The developed theory can be exploited to understand the data on
depinning of the DW from double $V$-shape constriction in
submicroscopic patterns of $\mathrm{SrRuO_{3}}$ \cite{Klein1}.
$\mathrm{SrRuO_{3}}$ is a metallic perovskite with orthorhombic
structure ($a=$5.53, $b=$5.57, $c=$7.82 $\mathrm{\AA}$) and an
itinerant ferromagnet with Curie temperature of $\sim\mathrm{150~K}$
and a saturated magnetization of $\sim\mathrm{1.4}\mu_{B}$ per
ruthenium. It shows so-called bad metal behavior at high
temperatures, but is a Fermi liquid at low temperatures.
$\mathrm{SrRuO_{3}}$ exhibits a positive Seebeck coefficient in the
wide range of the temperature from $\sim$0 K till $\sim$1000 K
\cite{Klein}, \cite{Maekawa}, manifesting the hole-like character of
the charge carriers. The samples are high-quality epitaxial thin
films of $\mathrm{SrRuO_{3}}$ grown by reactive electron beam
coevaporation on slightly miscut ($\sim \mathrm{0.2^{o})~SrTiO_{3}}$
substrates with the $\mathrm{[001]}$ and $\mathrm{[\bar{1}10]}$ axes
in the film plane. These films exhibit a large uniaxial
magnetocrystalline anisotropy (anisotropy field
$H_{a}\approx\mathrm{10}$ T at $T\to$ 0 K) with the easy axis tilted
out of the film ($\sim \mathrm{45^{o}}$) and with an in-plane
projection along $\mathrm{[\bar{1}10]}$. Consequently, the domain
magnetization is out of film plane, the Bloch DWs are parallel to
$\mathrm{[\bar{1}10]}$ axis and the orthorhombic anisotropy
(including the magnetostatic energy related to the shape anisotropy
of the sample) contributes to the structure of the DW. Due to the
large uniaxial magnetic anisotropy the DWs are relatively narrow
with temperature independent width of $\sim\mathrm{3~nm}$.

\begin{figure}[t]
  \includegraphics[width=0.4\textwidth]{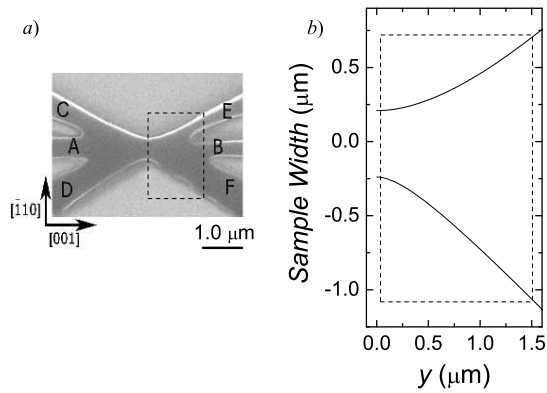}
\caption{ (a) Scanning electron microscope image of the patterned
sample \cite{Klein2} and (b) the shape function of the constricted
sample. Current pulses are injected between A and B. The average
magnetization is determined by measurement of extraordinary Hall
effect between C and D and between A and B. The rectangle bounded by
the dashed line shows the region of the sample where the DW
experiences a geometric pinning.} \label{Fig_4}
\end{figure}

The experimental setup is shown in Fig.~\ref{Fig_4} (a). The
measurements on the displacement of the DW driven by magnetic field
and spin-polarized current were performed on a high-quality
$\mathrm{375~\AA}$ thick film of $\mathrm{SrRuO_{3}}$ with the
resistivity ratio of $\mathrm{20}$
($\sim\mathrm{10}\mu\Omega~\mathrm{cm}$ at $\mathrm{4~K}$). The DW
initially located at the constriction (see, Fig.~\ref{Fig_4} (a))
was unpinned under the action of a magnetic field or a spin-polarized
current and moved in the positive direction of $y$-axis (parallel to
$\mathrm{[001]}$) towards the pair of leads EF. The magnetic state
of the sample in the region of contacts EF was monitored by
measuring of extraordinary Hall effect (EHE) proportional to the
average component of magnetization $M_{z}$ perpendicular to the
film plane ($xy$-plane). The final location of the DW at the leads
EF was deduced by the change of the sign of EHE followed by the
change of the magnetic state at EF. The experiment shows that the DW
displacement into a final position at the leads EF is achieved only
with a value of the magnetic field (current density) above a certain
threshold for the magnetic field or the current.

The shape of the sample [see Fig.~\ref{Fig_4} (b)] can
be approximated by a function
\begin{equation}\label{eq:Shape Function Klein}
\mathcal{G}(\eta)=\frac{(\eta/\zeta_{1})^{2}}{1+\eta/\zeta_{2}}+\frac{(\eta/\zeta_{3})^{2}}{1+\eta/\zeta_{4}}~,
\end{equation}
where $\zeta_{1}$=1.138 $\mu$m, $\zeta_{2}$=2.6 $\mu$m,
$\zeta_{3}$=0.561 $\mu$m, and $\zeta_{4}$=0.52 $\mu$m. The results of
numerical calculation show that a function
$G(\eta)=\partial_{\eta}\mathcal{G}(\eta)/[1+\mathcal{G}(\eta)]$ has
a maximum at $\eta=\zeta$=0.44 $\mu$m with $G(\zeta)$=1.1675
$\mu$m$^{-1}$. The assumption $\xi\ll\zeta$,  which is true assuming that
$\xi\sim \Delta\sim$ 1.5 nm, leaves two possibilities for the critical
value of magnetic field $H_{c}$ [see  Eq.~(\ref{eq:Depinning
Field})]:
\begin{equation}\label{eq:Depinning Field Klein}
H_{c}=\max\left\{ H_{\zeta}~;
H_{p}+\frac{1}{2}H_{a}\xi\Delta\left(\frac{1}{\zeta^{2}_{1}}+\frac{1}{\zeta^{2}_{3}}\right)\right\}~,
\end{equation}
where we  use
$\mathcal{G}(\eta)\approx\eta^{2}(\zeta_{1})^{-2}+\zeta_{1})^{-2}$
for $\eta\sim\xi$. The choice of $H_{c}$ in Eq.~(\ref{eq:Depinning
Field Klein}) depends on the critical value of magnetic field
corresponding to the geometrical pinning $H_{\zeta}$ given by
Eq.~(\ref{eq:Shape Critical Depinning Field}). Using the values of
$H_{a}\sim$10 T and $\Delta_{0}\sim$1.5 nm \cite{Klein1}, we obtain
$H_{\zeta}\sim$ 160 Oe, which is less than the highest value of the
depinning field $H_{c}=$ 500 Oe measured at the temperature $T=$ 40
K \cite{Klein1} . Therefore, we conclude that the  critical magnetic
field measured  in Ref.~\onlinecite{Klein1} is dominated by the
contribution of bulk pinning on defects, i.e., $H_{c}\approx H_{p}$,
in accordance with the conclusions of \cite{Klein1}. Evaluating  the
critical field $H_{c}$ (\ref{eq:Depinning Field Klein}), we have
neglected the small contribution of
$(\xi\Delta/2\zeta^{2})H_{\zeta}$, which is on the order of several
Oe. It follows from Eq.~(\ref{eq:Crititical Displacement}) that the
critical value of the DW displacement $\eta$ is $\xi\sim\Delta_{0}=$
1.5 nm $\ll\zeta=$ 0.44 $\mu$m, therefore the scenario illustrated
in Fig.~\ref{Fig_2} (b) is realized. After depinning when $H\ge
H_{\xi}$ and $\eta\ge\xi\sim\Delta_{0}=$ 1.5 nm the DW moves freely
(see Ref. \onlinecite{Tatara2}) till it reaches the leads EF
(Fig.~\ref{Fig_4} (a)).

In case of current driven DW motion the measured value of a
spin-polarized current corresponds to the arriving of the DW at the
leads EF (see Fig.~\ref{Fig_4} (a)). Both contributions to the
DW pinning resulting from the distribution of the defects and the
change of the sample shape can be evaluated from the data
\cite{Klein1} by use of Eq.~(\ref{eq:Depinning by Current 2}). To
calculate the value of the current predicted by the theory we
replace $\xi\partial_{\eta}f(\eta)$ in Eq.~(\ref{eq:Depinning by
Current 2}) by its maximum value:
$\max\{\xi\partial_{\eta}f(\eta)\}=$ 1,
\begin{equation}\label{eq:Depinning by Current Klien}
\frac{\beta U_{0}}{\gamma\Delta_{0}}-
H_{a}\Delta_{0}~\partial_{\eta}
\mathcal{G}(\eta)-[1+\mathcal{G}(\eta)] H_{p}=0~.
\end{equation}
Furthermore, we assume that the DW moves at the distance
$\eta\approx$ 1.5 $\mu$m till it is registered by observation of the
change of the sign of extraordinary Hall effect at the leads EF
(see, Fig. \ref{Fig_4}). With use of $H_{a}\approx$ 10 T,
$\Delta_{0}=$ 1.5 nm and the measured value of the depinning field
$H_{p}=$ 571 Oe at $T=$ 40 K \cite{Klein1} one can show that
$H_{a}\Delta_{0}~\partial_{\eta} \mathcal{G}(\eta)\approx$ 411 Oe
and $[1+\mathcal{G}(\eta)] H_{p}\approx$ 2250 Oe which gives a
relative contribution of the geometrical pinning $\sim$ 20 $\%$. The
measurement of the corresponding current density allows to evaluate
the parameter of the non-adiabaticity $\beta$. For
$\mathrm{SrRuO_{3}}$, the current density $J$ translates into the
velocity $U_{0}$ according to $U_{0}$[m/s]= 3.64 $\times$ 10$^{-10}
PJ$[A/m]. Substituting the measured value of the current density
$J=$ 5.8 $\times$ 10$^{10}$ A/m$^{2}$ into Eq.~(\ref{eq:Depinning by
Current Klien}) \cite{Klein1} and using the value of spin
polarization $P\approx$ 0.5 (see Ref. \onlinecite{Nadgorny}) we
obtain $\beta\approx$ 6. Being in accord with the conclusions of the
high efficiency of monitoring of the DW by spin-polarized current in
$\mathrm{SrRuO_{3}}$ \cite{Klein1}, such large value of $\beta$
cannot be explained by the contribution of the spin-relaxation
process, which gives the value of $\beta\sim\alpha\ll$1 (see Refs.
\onlinecite{Thiaville}, \onlinecite{Zhang}), but can be understood
due to the dominant role of the reflection of the charge carriers
from thin DWs in the framework of the theory developed in Ref.~
\onlinecite{Tatara1}.

Since bulk pinning on defects varies from a sample to sample, it is
unpractical to look for quantitative comparison of the theory and
the experiment in the case of predominantly bulk pinning. However,
if there are data for the critical magnetic field and the critical
current for the {\em same} sample, one may easily find from the
theory their ratio and compare it with the experimental values. As
it was shown by Tatara and Kohno \cite{Tatara1} the dynamics of the
abrupt DW in ideal plate-parallel sample is dominated by the
momentum transfer from the  charge current to the DW via reflection
of charge carriers from the DW. Since this momentum transfer
determines also the DW contribution to the resistance, in accordance
with Ref. \onlinecite{Tatara1} the parameter $\beta$ can be found
from the relation
\begin{equation}\label{eq:beta}
\frac{\beta
U_{0}}{\gamma\Delta_{0}}=\frac{enR_{w}\mathcal{A}}{2M_{s}}J~,
\end{equation}
where $n$ is a total charge carrier density and $R_{w}$ is the DW contribution to the
resistance. Substituting Eq.~(\ref{eq:beta}) into
Eq.~(\ref{eq:Depinning by Current 2}), we arrive at the equation
that defines the value of the current  density $J$ as a function of the
displacement of the DW in the presence of bulk and geometric pinning. As was pointed out in Sec.~\ref{sec:Depinning by spin-polarized current},  complete depinning of the DW from the potential produced by growing cross-section of the sample is impossible.However, one may introduce the critical current density  $J_c$ (determined at the constriction), at which the DW reaches the position where the DW is detected (the leads EF in   Fig.~\ref{Fig_4}). Then
\begin{equation}\label{eq:Current: Theory}
J_c=J_{c0}[1+\mathcal{G}(\eta)]\left[1+\frac{H_{a}}{H_{c}}\frac{\Delta_{0}\partial_{\eta}\mathcal{G}(\eta)}{1+\mathcal{G}(\eta)}\right]~,
\end{equation}
where $\eta$ is the distance between the sample constriction and the place of DW detection, and
\begin{equation}\label{eq:Tatara}
J_{c0}=\frac{2M_{s}H_{c}}{en\mathcal{A}_{0}R_{w}}
\end{equation}
 is the critical current density of depinning for the DW in an unconstricted
sample.

\begin{figure}[t]
  \includegraphics[width=0.4\textwidth]{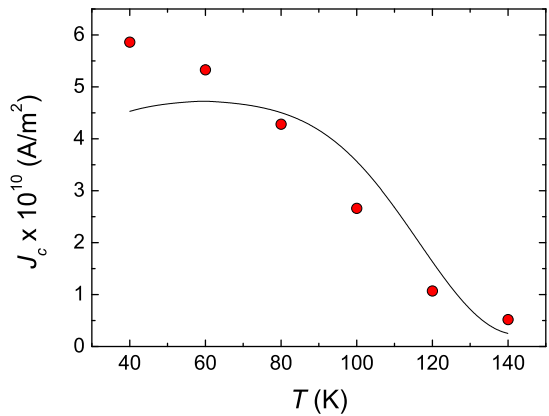}
\caption{\ Temperature dependence of the critical  current density
for constricted sample $\mathrm{SrRuO_{3}}$ (solid line) calculated
using experimental values of the critical magnetic field and taking
into account  the shape effect. The measured current  densities
(circles) are the data from Ref. \onlinecite{Klein1}.} \label{Fig_5}
\end{figure}

The temperature dependent saturation magnetization $M_{s}=M_{s}(T)$
and the DW resistance $R_{w}$ were measured in Refs.
\onlinecite{Klein3} and  \onlinecite{Klein1} respectively. According
to Ref.~\onlinecite{Klein1} the carrier density $n$ is about 1.6
$\times$ 10$^{28} $ [1/m$^3$]. This offers an opportunity to compare
the developed theory with the data \cite{Klein1}. Calculating the
critical current density $J_c$ with help of Eq.~(\ref{eq:Current:
Theory}), we assume that the relative contribution of the geometric
pinning [the term $\propto H_a$ in Eq.~(\ref{eq:Current: Theory})]
is temperature independent and equals to the value of $\sim$ 20 $\%$
calculated at $T=$ 40 K. Then  Eq.~(\ref{eq:Current: Theory}) yields
$J_c\approx$5$J_{c0}$. Figure ~\ref{Fig_5}  shows the experimentally
determined critical current density (cicles) together with the
prediction of the theory using the experimentally found critical
magnetic fields and taking into account the shape dependent effect
(geometric pinning). The figure illustrates a satisfactory agreement
between the experiment and the theory.

It is important to note that the sign of the ratio between the
current and the DW displacement depends on the sign of charge
carriers. The relative sign of the current and the displacement in
the experiment gives evidence that charge carriers are holes. This
agrees with the experiment on the Seebeck effect \cite{Klein}.

\section{\label{sec:Conclusions}Conclusions}

We investigated pinning of a domain wall by potentials produced by
bulk defects and the sample shape. The process of depinning by an
external magnetic field and by a spin-polarized current was
analyzed. The shape-dependent pinning potential (geometric pinning)
can essentially affect the process of depinning and may even make
complete depinning  by the spin-polarized current impossible. Though
the absolute values of the critical magnetic fields and the critical
currents, at which depinning occurs, are sample dependent and
difficult for theoretical prediction, their ratio must be sample
independent \cite{Tatara1} and allows reliable comparison of the
theory and the experiment. We performed this comparison and found a
satisfactory agreement.

\section*{Aknowledgments}

We appreciate the useful discussions with L. Klein. This work has
been supported by the grant of the Israel Academy of Sciences and
Humanities and grand of the Israel Scientific Foundation (Grand No.
499/07).

  \end{document}